\documentclass[]{spie}  
\usepackage[]{graphicx}

\def\3he{$^3{\rm He}$}
\def\4he{$^4{\rm He}$}

\newcommand{\wmap}{{\it WMAP}}
\newcommand{\planck}{{\it Planck}}
\newcommand{\herschel}{{\it Herschel}}
\newcommand{\blast}{BLAST}
\newcommand{\blastpol}{BLAST-pol}

\newcommand{\scubatwo}{SCUBA-2}

\def\deg{$^\circ $}
\newcommand{\arcmin}{\hbox{$^\prime$}}               
\newcommand{\arcsec}{\hbox{$^{\prime\prime}$}}       
\newcommand{\mic}{$\mu$m}                            

\title{The Balloon-borne Large-Aperture Submillimeter Telescope for polarization: \blast-pol} 

\author{
        G.~Marsden,\supit{a}
        P.~A.~R.~Ade,\supit{b}
        S.~Benton,\supit{c}
        J.~J.~Bock,\supit{d,e}
        E.~L.~Chapin,\supit{a}
        J.~Chung,\supit{a,c}
        M.~J.~Devlin,\supit{f}
        S.~Dicker,\supit{f}
        L.~Fissel,\supit{c}
        M.~Griffin,\supit{b}
        J.~O.~Gundersen,\supit{g}
        M.~Halpern,\supit{a}
        P.~C.~Hargrave,\supit{b}
        D.~H.~Hughes,\supit{h}
        J.~Klein,\supit{f}
        A.~Korotkov,\supit{i}
        C.~J.~MacTavish,\supit{c}
        P.~G.~Martin,\supit{j,k}
        T.~G.~Martin,\supit{c}
	T.~G.~Matthews,\supit{l}
        P.~Mauskopf,\supit{b}
	L.~Moncelsi,\supit{b}
        C.~B.~Netterfield,\supit{c,k}
	G.~Novak,\supit{l}
        E.~Pascale,\supit{b}
	L.~Olmi,\supit{m,n}
        G.~Patanchon,\supit{a,o}
        M.~Rex,\supit{f}
	G.~Savini,\supit{b}
        D.~Scott,\supit{a}
        C.~Semisch,\supit{f}
        N.~Thomas,\supit{g}
        M.~D.~P.~Truch,\supit{f}
        C.~Tucker,\supit{b}
        G.~S.~Tucker,\supit{i}
        M.~P.~Viero,\supit{k}
	D.~Ward-Thompson,\supit{b}
        D.~V.~Wiebe\supit{c}
\skiplinehalf
\supit{a}{Department of Physics \& Astronomy, University of British Columbia, 6224 Agricultural Road, Vancouver, BC V6T~1Z1, Canada};\\
\supit{b}{Department of Physics \& Astronomy, Cardiff University, 5 The Parade, Cardiff, CF24~3AA, UK};\\
\supit{c}{Department of Physics, University of Toronto, 60 St.~George Street, Toronto, ON M5S~1A7, Canada};\\
\supit{d}{Jet Propulsion Laboratory, Pasadena, CA 91109-8099};\\
\supit{e}{Observational Cosmology, MS 59-33, California Institute of Technology, Pasadena, CA 91125};\\
\supit{f}{Department of Physics and Astronomy, University of Pennsylvania, 209 South 33rd Street, Philadelphia, PA 19104};\\
\supit{g}{Department of Physics, University of Miami, 1320 Campo Sano Drive, Coral Gables, FL 33146};\\
\supit{h}{Instituto Nacional de Astrof\'isica \'Optica y Electr\'onica (INAOE), Aptdo. Postal 51 y 72000 Puebla, Mexico};\\
\supit{i}{Department of Physics, Brown University, 182 Hope Street, Providence, RI 02912};\\
\supit{j}{Canadian Institute for Theoretical Astrophysics, University of Toronto, 60 St.~George Street, Toronto, ON M5S~3H8, Canada};\\
\supit{k}{Department of Astronomy \& Astrophysics, University of Toronto, 50 St.~George Street, Toronto, ON M5S~3H4, Canada};\\
\supit{l}{Department of Physics and Astronomy, Northwestern University, 2145 Sheridan Road, Evanston, IL 60208-3112};\\
\supit{m}{Istituto di Radioastronomia, Largo E. Fermi 5, I-50125, Firenze, Italy};\\
\supit{n}{University of Puerto Rico, Rio Piedras Campus, Physics Dept., Box 23343, UPR station, San Juan, Puerto Rico};\\
\supit{o}{Laboratoire APC, 10, rue Alice Domon et L{\'e}onie Duquet 75205 Paris, France};
}

\authorinfo{Further author information: (Send correspondence to
  G. Marsden)\\G. Marsden: E-mail: gmarsden@phas.ubc.ca, Telephone: 604-822-6709}

\begin{document} 
  \maketitle

\begin{abstract}
The Balloon-borne Large Aperture Submillimeter Telescope (\blast) is a
sub-orbital experiment designed to study the process of star formation
in local galaxies (including the Milky Way) and in galaxies at
cosmological distances.  Using a 2\,m Cassegrain telescope,
\blast\ images the sky onto a focal plane, which consists of 270
bolometric detectors split between three arrays, observing
simultaneously in 30\% wide bands, centered at 250,
350, and $500\,\mu$m\@.  The diffraction-limited optical
system provides a resolution of $30\arcsec$ at $250\,\mu$m\@.  The
pointing system enables raster-like scans with a positional accuracy
of $\sim 30\arcsec$, reconstructed to better than 5\arcsec\ rms in
post-flight analysis\@.  \blast\ had two successful flights, from the
Arctic in 2005, and from Antarctica in 2006, which provided the first
high-resolution and large-area ($\sim$\,0.8\,--\,200 deg$^2$)
submillimeter surveys at these wavelengths.  As a pathfinder for the
SPIRE instrument on \herschel, \blast\ shares with the ESA satellite
similar focal plane technology and scientific motivation. A third
flight in 2009 will see the instrument modified to be
polarization-sensitive (\blastpol). With its unprecedented mapping
speed and resolution, \blastpol\ will provide insights into Galactic
star-forming nurseries, and give the necessary link between the
larger, coarse resolution surveys and the narrow, resolved
observations of star-forming structures from space and ground based
instruments being commissioned in the next 5 years.
\end{abstract}



\section{INTRODUCTION}
\label{sec:intro}  
The Balloon-borne Large Aperture Submillimeter Telescope (\blast) is a
stratospheric 2\,m telescope which observes the sky with bolometric
detectors operating in three 30\% wide bands at 250, 350, and
500$\,\mu$m. The diffraction-limited optics are designed to provide
\blast\ with a resolution of 30\arcsec, 42\arcsec, and 60\arcsec\ at the
three wavebands, respectively. The detectors and cold optics are
adapted from those to be used on the SPIRE instrument on
\herschel\cite{grif03}.

\blast\ addresses important Galactic and cosmological questions
regarding the formation and evolution of stars, galaxies and
clusters\cite{dev04} by providing the first large-area
(0.8--200\,deg$^2$) surveys of unique spectral-coverage, angular
resolution and sensitivity.  The primary scientific goals of
\blast\ are: (1) to conduct confusion-limited and shallower wide-area
extragalactic surveys to constrain the redshift-distribution, star
formation history, and evolution of optically-obscured luminous
galaxies by measuring photometric redshifts (derived from the \blast\
colors and complementary data\cite{hughes02}); (2) to study the
spatial clustering of this population; (3) to improve our
understanding of the earliest stages of star formation by determining
the physical properties and mass-function of cold pre-stellar cores
and the efficiency of star formation within different Galactic
environments; and (4) to investigate the nature and structure of the
interstellar medium by making the highest resolution maps to date of
diffuse Galactic emission at these wavelengths.

\blast\ has had two Long Duration Balloon (LDB) flights. The first was
a 4-day flight from Kiruna, Sweden in June of 2005 (BLAST05). The
primary mirror was slightly damaged on launch or ascent preventing
sensitive extragalactic observations. However, we were able to observe
multiple Galactic targets resulting in a successful series of
observations\cite{truch07,pascale07,patanchon07,chapin07}.  The
instrument was quickly repaired and flew again from Antarctica in
December 2006 with an 11-day flight (BLAST06). This time the entire
instrument worked perfectly. We obtained multiple deep large-area
Galactic and extragalactic maps.

The termination of the flight in Antarctica resulted in the payload
being dragged 200\,km over the course of 24\,hr. Fortunately, the hard
drives containing the flight data were recovered. We were also able to
recover the detectors, the only irreplaceable component, as well as
the receiver and the primary and secondary mirrors. This allowed us to
continue with our long-term plans for the program. The reconstruction
of the instrument is well-underway and it is expected to be completed
by mid 2009.

The simple transformation of \blast\ into \blastpol\ will make it a
uniquely sensitive experiment for mapping polarized emission from
Galactic dust. \blastpol\ will be a powerful instrument for studying
the hotly debated question of whether it is magnetic fields or
turbulence that controls star formation. In particular, we will be
able to map large-scale fields over an entire Giant Molecular Cloud
(GMC) with enough angular resolution to trace the fields into cloud
cores and dense filaments. Scales accessible to \blastpol\ lie in
between those soon accessible to \planck\ (5\arcmin\ resolution),
\scubatwo\ (7\arcsec\ at 450\,\mic) and ALMA (sub-arcsecond).

\section{Science Goals}

\subsection{Results}
\blast\ was designed to conduct confusion-limited and wide-area
extragalactic and Galactic surveys at submillimeter (submm)
wavelengths from an LDB platform. \blast's wavelengths are impossible
or very difficult to observe from even the best ground-based telescope
sites.  In the Kiruna flight we were able to observe multiple Galactic
targets resulting in a successful series of observations (Truch et
al. 2008; Pascale et al. 2008; Patanchon et al. 2007; Chapin et
al. 2007; see also Fig.~\ref{fig:mag1}). The first results from
BLAST06 are in preparation. At least 10 more publications are planned
before the dataset will be released to the community. These results,
important on their own, will also guide the SPIRE team in planning
observations and making the best use of this mission. The team is now
in the process to add polarization capability to the existing focal
plane: the \blastpol\ experiment.

\subsection{Mapping the Galactic Large-Scale Magnetic Fields}
\begin{figure}
  \begin{center}
  \begin{tabular}{ccc}
    \includegraphics[height=5cm, keepaspectratio]{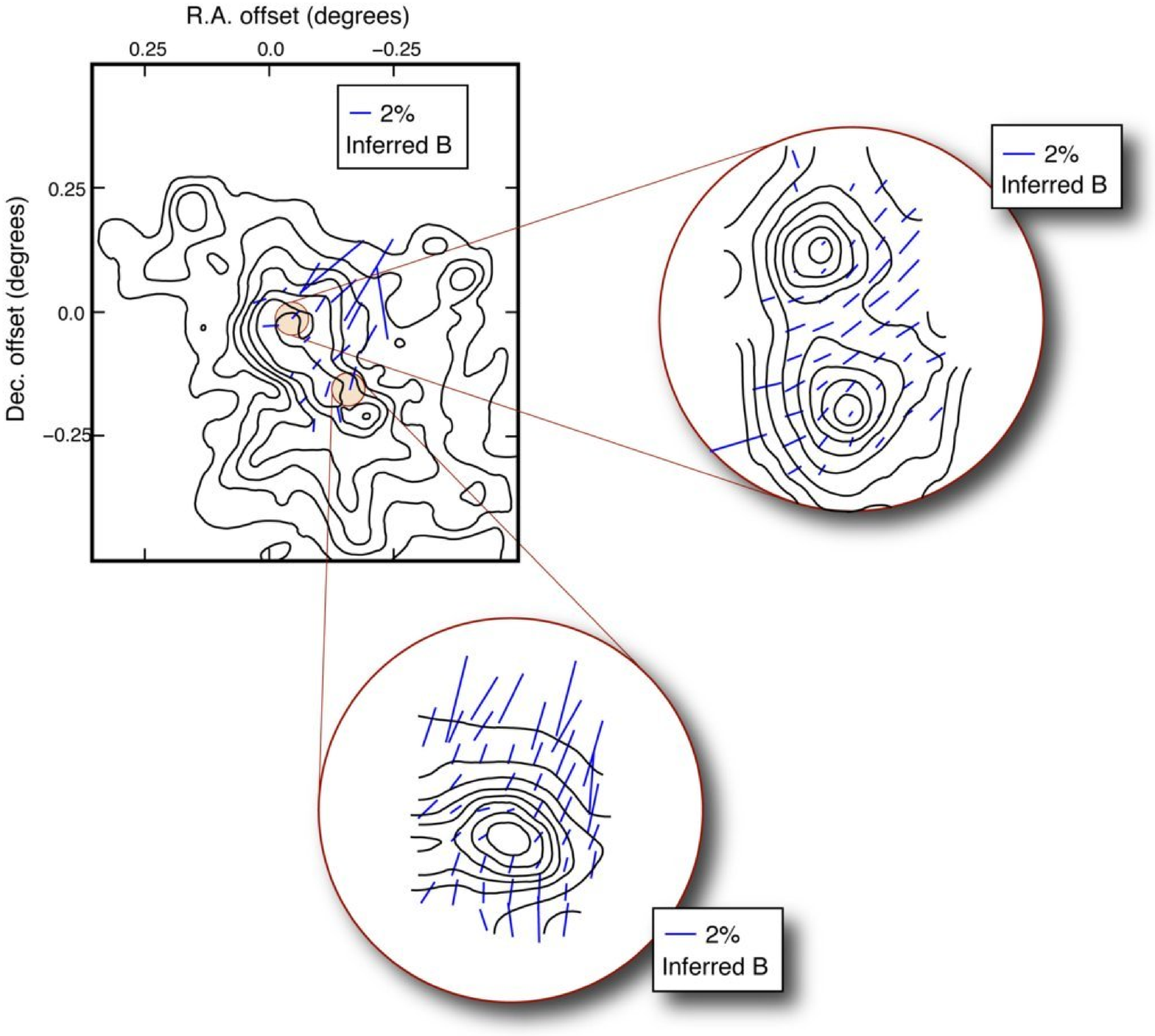}&&\includegraphics[height=5cm, keepaspectratio]{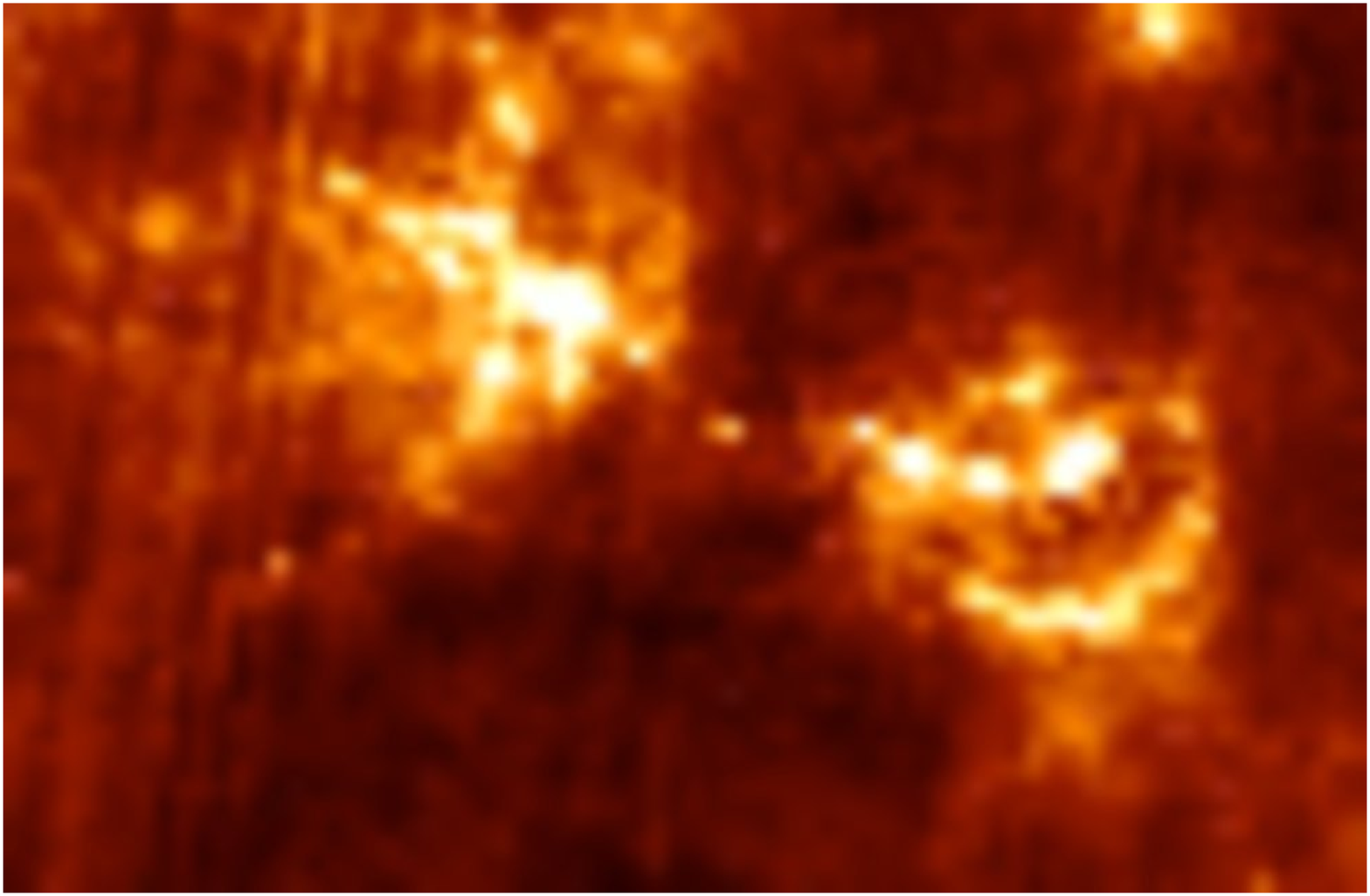}
  \end{tabular}
  \end{center}
  \caption[example] { \label{fig:mag1} {\em Left panel:\/}
    submillimeter polarimetry of NGC 6334, a massive GMC at a distance
    of 1700 pc. The pseudo-vectors in the map at upper left were
    obtained using SPARO at South Pole, and the ``blowup'' at right
    and bottom are from Hertz/CSO on Mauna Kea. The orientation of
    each pseudo-vector shows inferred field direction, and the length
    of each pseudo-vector is proportional to the degree of
    polarization. With \blast-pol, we will map scores of such GMCs,
    obtaining up to 1000 pseudo-vectors per cloud. {\em Right
      panel:\/} A dust emission map at 250\,\mic\ of a star forming
    region in the Galactic plane towards the constellation Cygnus
    obtained by \blast\ during the 2005 flight. }
   \end{figure}

Despite many advances, fundamental questions about star formation
remain unanswered\cite{mckee07}. Is star formation regulated by
magnetic fields or by turbulence? Do molecular clouds and their
substructures (cores, filaments, and clumps) have lifetimes exceeding
their turbulent crossing times? Magnetic fields in molecular clouds
are notoriously difficult to observe\cite{crutch04,whittet08}. A
promising method for probing these fields is far-IR/submm
polarimetry\cite{hildebrand00, ward00}. \blastpol\ will be the first
submm polarimeter having both (a) sufficient mapping speed to trace
fields across entire clouds, for a statistically significant sample of
GMCs, and (b) sufficient spatial resolution to follow the cloud fields
into the dense cores. \blastpol\ will map polarized emission from Av =
100\,mag out to Av = 4\,mag, yielding $\sim 1000$ polarization
detections per cloud, for scores of clouds. No other instrument will
have this capability. These data will allow the first detailed
comparisons between observed GMC field maps and synthetic GMC field
maps --- the latter derived from numerical turbulence
simulations\cite{ostriker01}. This will allow us to make detailed
observational tests of theoretical magneto hydrodynamic (MHD)
models. Recent observations made at South Pole show that the extended
submm emission from GMCs is indeed polarized\cite{li06}
(Fig.~\ref{fig:mag1}). We will use \blastpol\ observations
(Fig.~\ref{fig:mag1}) to probe magnetic field patterns on intermediate
scales. This will allow us to tackle quite detailed questions such as
the following 3 examples:
\begin{enumerate}
\item Is core morphology determined by large-scale fields? Jones \&
  Basu\cite{jones02} argue that observations support the predominance
  of oblate cores in molecular clouds, as predicted by
  magnetically-regulated models\cite{mous99, allen03}. Such models
  also predict that each core is embedded in a large-scale cloud field
  running parallel to the core minor axis. Submm polarimetry of
  quiescent cores\cite{ward00} questioned this prediction by finding
  in SCUBA data that magnetic fields were misaligned, more consistent
  with turbulent magnetic models. However, SCUBA (and \scubatwo) cannot
  trace the fields into lower density regions, and what is really
  needed are observations tracing core fields out into the surrounding
  lower-density cloud regions.  \blastpol\ will provide these.
\item Does filamentary structure have a magnetic origin?  Faint,
  tenuous $^{12}$CO filaments observed in
  Taurus\cite{heyer08,goldsmith08} closely follow the optical
  polarimetry vectors, indicating a possible magnetic origin.
  However, denser filaments in Taurus and other clouds show no
  preferred orientation with respect to the nearby stellar polarimetry
  vectors\cite{goodman90}. This may reflect a non-magnetic origin for
  the denser filaments, or may simply reflect the inadequacies of
  optical/near-IR polarimetry for tracing fields in shielded
  regions\cite{whittet08,cho05}. It is therefore important to map
  dense filaments using submm polarimetry to answer the question about
  the relation between fields and structure.
\item What is the field strength, and how does it vary from cloud to
  cloud?  The Chandrasekhar-Fermi (CF) technique relates dispersion in
  polarization angle to field strength\cite{chandra53,zweibel90}.  For
  molecular cloud cores, CF field strength estimates have been
  obtained from submm data, and the results are in rough agreement
  with Zeeman observations\cite{crutch04}.  Numerical turbulence
  simulations have been used to calibrate the CF technique for
  molecular clouds\cite{ostriker01,pelkonen07,falceta08}. \blastpol\
  will determine the power spectrum of the angle dispersion over a
  wide range of spatial scales and will also measure 3-color
  polarization spectra and their spatial
  variations\cite{bethell07,vaillancourt08}.  These observed
  characteristics will provide new constraints for the turbulence
  models that should lead to improved models and better CF
  calibration. Our observations will probe the dependence of field
  strength on cloud age, cloud location, and cloud mass.
\end{enumerate}

\subsection{Polarized Galactic Dust Foregrounds}
The search for the primordial gravity waves (GWs) by detecting the
B-mode polarization of the Cosmic Micronwave Background
(CMB)\cite{hu97,kamionkowski97,zaldarriaga97} requires a
control of instrumental and astrophysical systematics at a level which
allows the detection of this faint signal.  Astrophysical foregrounds
might ultimately be a limiting factor, as the the cosmological B-mode
intensity has currently only upper limits and a solid theoretical
background is missing to make any realistic prediction. The
\wmap\ polarization observations\cite{page07} clearly demonstrate that
polarized galactic foregrounds, even in the range of frequencies where
their emission is close to a minimum, are likely to be comparable or
larger than the B-mode CMB polarization anisotropy at large angular
scales ($\ell < 100$) and intermediate Galactic latitudes. This is a
situation unlike the temperature anisotropy where most of the sky is
dominated by CMB, and it requires significant foreground emission
modeling in order to extract the underlying CMB polarization
signal. Most of the polarized galactic foreground that \wmap\ observed
is due to synchrotron emission. At the highest \wmap\ frequency band
(94\,GHz), there is evidence for polarized dust emission. This is
potentially problematic for the many higher-frequency bolometric CMB
polarization experiments (EBEX, Spider, Clover, QUAD, BICEP, SPUD,
\planck-HFI, and Polarbear) since the dust is expected to rise
significantly in frequency as predicted by models based on the
observations at 100$\,\mu$m. In addition to \wmap\ the ARCHEOPS
balloon mission measured Galactic dust polarization on scales of
15\arcmin\ at submm wavelengths. ARCHEOPS results\cite{ben04} have
shown that radiation from diffuse Galactic dust is polarized (as
expected) at a $3-5$\% level, but in some regions the polarization is
as high as 10\%. While the measurements were made near the Galactic
plane, they point to the need for more information at higher
latitudes, with higher sensitivity, better angular resolution, and at
multiple wavelengths. Despite the \wmap\ and Archeops results, no
information exists for any region of the sky at the accuracy required
for a B-mode signal detection, and very little information exists at
frequencies relevant to CMB science. There are also puzzles in dust
polarimetry that must be resolved if we are to make a robust detection
of the primordial GWs. For instance, polarimetry of bright molecular
cloud cores at 350$\,\mu$m and 1\,mm has reveled the surprising result
that the polarized fraction at 1\,mm is larger by a factor of 2
compared to that at 350$\,\mu$m\cite{hildebrand00}. The
\planck\ satellite will provide a unique dataset by mapping the sky
with polarization sensitive bolometers in bands centered from
850$\,\mu$m to 3\,mm. The 850$\,\mu$m band will provide an
unparalleled dataset for studying polarized dust. In addition to
\planck, balloon-borne and other ground based instruments employ
anywhere 2 to 6 frequency bands to assist in discriminating against
foreground contamination. But the ability to separate the foreground
dust emission to the necessary level to detect B-mode polarization
depends from the complexity of the polarized dust emission.

\begin{figure}
  \begin{center}
  \begin{tabular}{c}
    \includegraphics[width=0.4\linewidth, angle=-90,keepaspectratio]{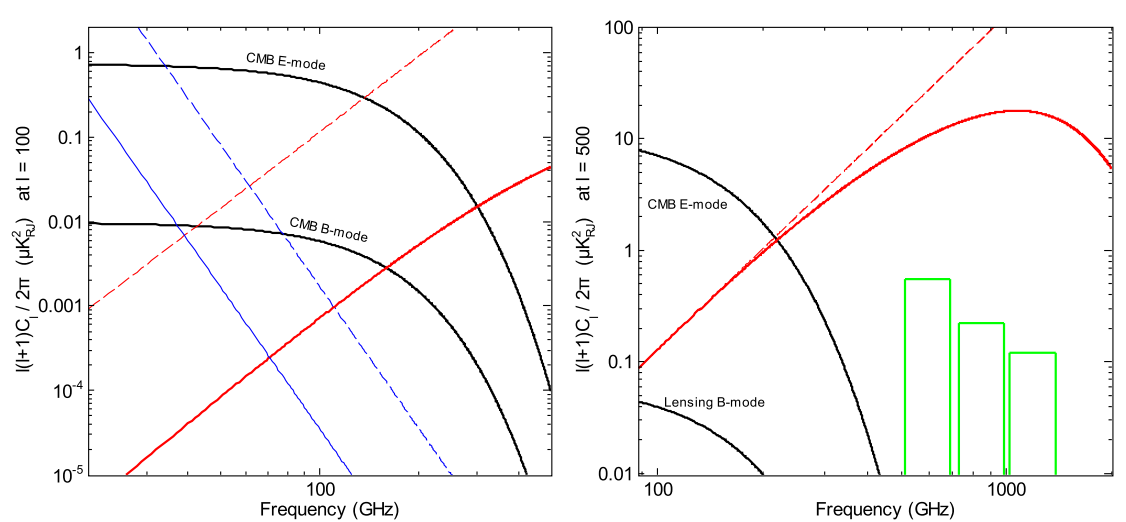}
  \end{tabular}
  \end{center}
  \caption[example] { \label{fig:foregrounds} {\it Left panel:\/} The
    spectrum of B-mode foregrounds from Galactic dust (red) and
    synchrotron (blue) expected outside \wmap's P06 mask\cite{page07}
    (dashed) and from a selected sky region (solid), and the E and
    B-mode CMB (black, solid). All at $\ell = 100$ and in antenna
    temperature. This dust model assumes model 8 from Finkbeiner et
    al. (1999)\cite{fink99}. {\it Right panel:\/} A similar plot at $\ell
    =500$. The green bars show the \blastpol\ noise at a resolution of
    8\arcmin\ from a 48\,hr observation of 10 deg$^2$ region. At these
    angular scales, the CMB signal is dominated by the lensing-B. The
    Galactic dust signal, normalized to \wmap, peaks at the
    \blastpol\ frequencies. (The CMB models assume a standard
    $\Lambda$CDM cosmology with r = 0.1.)  }
   \end{figure}

Figure~\ref{fig:foregrounds} shows the expected polarized dust
emission as measured by the 3\,yr \wmap\ data\cite{page07} and the level
of foregrounds predicted in a selected $\sim 200$\,deg$^2$ sky region
($l = 258$\deg, $b = -46$\deg) accessible by sub-orbital and ground
based instruments.  Amazingly, the dust signal extrapolated to the CMB
frequencies are comparable to the r\,=\,0.1 B-mode signal level that
is being probed by the current generation of experiments. For this
reason, it is imperative that we begin to understand high-latitude
polarized dust emission. Given these rather small signal levels,
\blastpol\ will not be able to conduct large surveys that could be
used as templates for lower frequency CMB polarization observations.
A 48-hr survey over a 5 deg$^2$ sky region
(Fig.~~\ref{fig:foregrounds}, right panel) will enable \blastpol\ to
constrain dust properties difficult to measure at lower
frequencies. In particular, the determination of dust temperature and
the degree of polarization will be highly degenerate as measured by
all CMB experiments in the Rayleigh-Jeans tail of the dust spectrum.
The ability of \blastpol\ to extract these parameters will be
straightforward since the measurements are near the peak of the dust
spectrum.

The first of our two maps will be aimed in a very low dust emission
region, comparable to 1 MJy/sr at 100$\,\mu$m, that is ideally
coincident with a deep CMB polarization measurement. The other will
occur in a mid-latitude region with a higher dust brightness that will
almost certainly be sampled by \planck. The corresponding higher signal
will enable parameter extraction without the aid of lower frequency
CMB polarization measurements.

When these measurements are combined, it will enable the
characterization of the spatial variation of the polarization
percentage, the dust temperature, and the dust emissivity. While the
maps will be smoothed to 8\arcmin\ pixels, we will retain the ability
to see any fine-scale, higher-level polarization signals to help the
planning of future ground-based, balloon-borne and space experiments.

\section{INSTRUMENT}
\subsection{Optical Design}

\begin{figure}
  \begin{center}
  \begin{tabular}{c}
    \includegraphics[height=5cm, keepaspectratio]{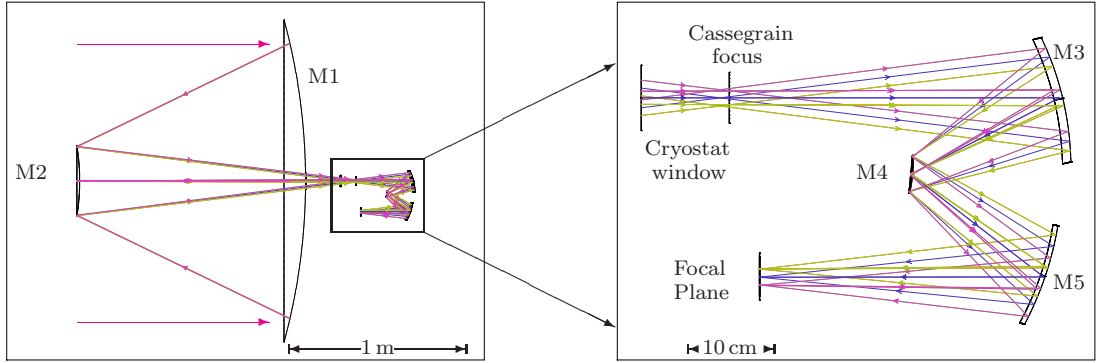}
  \end{tabular}
  \end{center}
  \caption[example] { \label{fig:optical} The optical layout of the
    \blast05 telescope and receiver is shown on the left and the 1.5K
    optics, located within the cryostat, are shown in expanded view on
    the right.  The image of the sky formed at the input aperture is
    re-imaged onto the bolometer detector array at the focal
    plane. The mirror M4 serves as a Lyot stop defining the
    illumination of the primary mirror for each element of the
    bolometer array. The three wavelength bands are separated by a
    pair of dichroic beamsplitters (not shown) which are arranged in a
    direction perpendicular to the plane, between M5 and the focal
    plane.  }
   \end{figure}

The \blast\ gondola incorporates a 1.9\,m primary mirror and a 40\,cm
diameter correcting secondary mirror giving diffraction-limited
performance over a $6.5\arcmin \times 13\arcmin$ FOV at the telescope
focus at $\lambda$ = 250\,$\mu$m. A diagram of the optical system is
given in Fig.~\ref{fig:optical}.  The estimated antenna efficiency is
$> 80$\%, determined by a combination of the rms surface roughness of
the primary and the quality of the re-imaging optics.

A carbon fiber spherical primary mirror was used for the Sweden
flight, but was damaged during the flight. For the Antarctic flight we
used a parabolic aluminum mirror in a Ritchie-Chreti\'en
configuration. This mirror, used during the 2003 test flight, was
resurfaced to an surface accuracy of $ < 4\,\mu$m rms by the Precision
Engineering Group at Lawrence Livermore National Laboratory.

Temperatures are not constant during the flight and thermal expansion
causes variation in the curvature radii of the optical sourfaces.  The
relative distance between the primary and the secondary mirrors has a
tolerance of one wavelength before significant image degradation is
introduced. Thermal modeling indicated that diurnal temperature
fluctuations at balloon altitudes (as high as 10\deg C) causes
requires a correction in the relative distance between the primary and
the secondary mirror of $50\,\mu$m/\deg C.

To compensate, we move the secondary using three stepper motor actuators
to correct for the focus position.  These actuators also provide tip/tilt
capability for initial alignment.

\begin{figure}
  \begin{center}
  \begin{tabular}{c}
    \includegraphics[height=5cm, keepaspectratio]{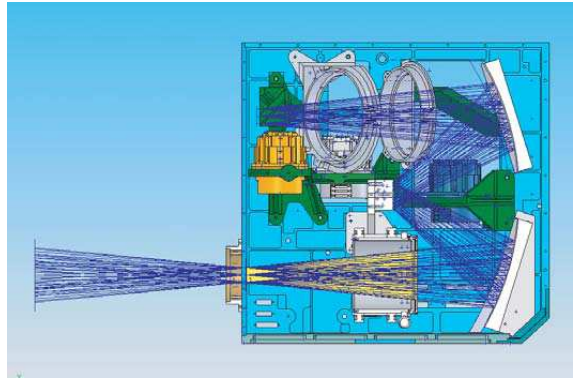}
  \end{tabular}
  \end{center}
  \caption[example]
   { \label{fig:opticsbox} 
	A cutaway view of the \blast\ optics
box. The light enters from the lower left and is reimaged
onto the arrays. Dichroic filters split the
beam into each of the arrays. For \blast-pol,
a modulating cold half-wave plate will be placed
inside the optics box and polarizing grids will be
mounted in front of each of the BDAs. The motorized
mechanism for rotating the waveplate will
be located outside the optics box.
   }
   \end{figure}

In the configuration used for the Antarctic campaign, radiation from
the telescope is re-imaged onto the focal plane by three spherical
mirrors arranged in an Offner-relay configuration and mounted in a
1.5\,K cooled optics box (Figs.~\ref{fig:optical} and
\ref{fig:opticsbox}).  A more complete discussion of the optics for
BLAST05 can be found in Olmi (2002)\cite{Olmi02}.

\subsection{Detectors}
The \blast\ focal plane consists of arrays of 149, 88 and 43 detectors
at 250, 350, and 500\,$\mu$m, respectively. The detectors are
silicon-nitride micromesh (“spider-web”) bolometric detectors coupled
with 2f$\lambda$ feedhorn arrays. The entire bolometer/detector array
(BDA) is based on the SPIRE instrument detectors\cite{bock98, row03}.

The sensitivity of the detectors is limited by photon shot-noise from
the telescope.  The total emissivity for the warm optics of $\sim$ 6\%
is dominated by blockage from the secondary mirror and supports. We
estimate the optical efficiency of the cold filters and optics to be
$\eta_{\rm opt} \ge 0.3$.  The detectors are read out with an
AC-biased differential circuit. The data acquisition electronics
demodulate the detector signals to provide noise stability to low
frequencies ($<$ 30 mHz), which allows the sky to be observed in a
slowly-scanned mode. Slow scanning is preferable to a mechanical
chopper for mapping large regions of sky to the confusion limit.  The
data are collected using a high-speed, flexible, 22-bit data
acquisition system developed at the University of Toronto. The system
can synchronously sample up to 600 channels at any rate up to 4
kHz. Each channel consists of a buffered input and a sigma-delta
analog to digital converter.  The output from 24 channels are then
processed by an Altera programmable logic device (PLD) which digitally
anti-alias filters and demodulates each input. The results are stored
to disk.

\subsection{Polarimetry}
\begin{figure}
  \begin{center}
  \begin{tabular}{c}
    \includegraphics[height=5cm, keepaspectratio]{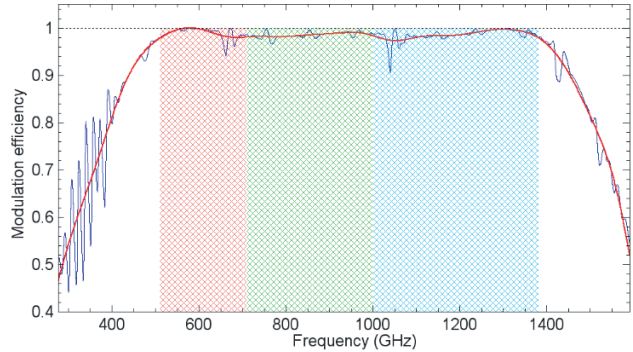}
  \end{tabular}
  \end{center}
  \caption[example]
   { \label{fig:modeff} 
	The modeled performance of the antireflection
coated 5-plate 
design of
the \blast\ half-wave plate. The diameter will
be 9 cm. A similar design has already been built
and tested for \scubatwo\ (20 cm diameter). The
optical efficiency ranges between 85 and 96\% depending
on the type of anti-reflection coating. We
will build several types and use the one with the
best results.
   }
   \end{figure}

Adding polarimetry to the \blast\ focal plane is relatively
straightforward compared to the technical challenges that have been
addressed in building the whole instrument. The design incorporates an
Achromatic Half Wave Plate (AHWP) located inside the optics box as
indicated in Fig.~\ref{fig:opticsbox}. It will be mounted inside the
optics box, far from the Cassegrain focus to minimize modulation of
imperfections, while allowing it to operate and dissipate heat to the
4\,K stage. The 5 layer AHWP will have an optical efficiency between
85 and 96\% depending upon the anti-reflection coating and material
used. Options currently under investigation include quartz and
sapphire. Each of these presents different challenges and we will
chose the solution which maximizes the ratio between optical
performances to technical challenge. The predicted modulation
efficiency across the \blast\ bands is given in Fig.~\ref{fig:modeff}.
The waveplate will be operated in either a stepped mode or
continuously rotated. The stepped mode solution implements a similar
mechanism to the one used by the SPARO instrument\cite{renb04}. The
continuous rotation will be obtained using magnetic levitating
bearings similar to the mechanism developed for Clover. Rotation rate
will range from 1 to 5\,Hz resulting in a polarization modulation of 4
to 20\,Hz. The two mechanisms, driven by a 300\,K external motor, are
interchangeable in order to allow maximum flexibility. The solution
and rotation frequency which will be adopted for the flight will
depend from the response of the instrument to the different modulation
methodologies.

A polarizing photolithographed grid in front of each feed-horn array
will be patterned to alternate the polarization angle from horn-to-horn
along the scan direction. \blast\ has the demonstrated pointing
stability necessary to control the scan direction and pitch to have a
point on the sky pass from horn-to-horn along a row of the array. The
alternating grid will modulate the polarization at beam-crossing
timescales even if stepping system for the half-wave plate fails. The
placement of the grid is similar to the method used by
Boomerang\cite{masi06}.

\subsection{Cryogenics}
The receiver consists of an optical cavity inside a long hold-time
liquid-nitrogen and liquid-helium cryostat. Both the nitrogen and
helium are maintained at slightly more than atmospheric pressure
during the flight to minimize loss due to pressure drop at altitude.
A \3he refrigerator maintains the detectors at 300\,mK during
flight. The self-contained, recycling refrigerator can maintain a base
temperature of 280\,mK with 30\,$\mu$W of cooling power for 4 days. It
can be recycled within 2\,hr. The \3he refrigerator uses a pumped \4he
pot at $\sim 1$\,K for cycling and to increase the hold time of the
system. The pumped pot maintains 1\,K with 20\,mW of cooling power
with outside pressure 15 Torr or less. The entire optics box
containing the re-imaging optics is also cooled to 1 K.

\subsection{Gondola}
The \blast\ gondola provides a pointed platform for the telescope and
the attachment point to the balloon flight train.  The instrument is
mounted on an alt-azimuth aluminum frame (the gondola) made of an
outer frame (OF) providing azimuth directionality and holding an inner
frame for elevation pointing. The OF is hung from the $1.1 \times
10^6$\,m$^3$ helium balloon, provided by NASA's Columbia Scientific
Balloon Facility, through a steel cable and parachute. Control
computers and telemetric systems are mounted on the OF. Data are
stored on these computers and partially transmitted through satellite
links to the ground station. Sun shields, made of aluminized mylar
panels, keep sunlight from heating the telescope, and are mounted on
the OF.  The IF is built to house the mirror, the receiver, its
readout electronics and the pointing sensors, all rigidly mounted with
respect to each other to ensure that mechanical alignment is
maintained throughout the flight.

The telescope is controlled in azimuth by a 1.5\,m reaction wheel,
made of 7.6\,cm thick aluminum honeycomb, and 48 0.9\,kg brass
disks mounted around the perimeter, to maximize the ratio of moment of
inertia to mass. The reaction wheel is mounted at the center of the
outer frame with its rotation axis going through the pivot. Torquing
the gondola against the reaction wheel controls azimuth pointing. The
elevation of the inner frame is controlled by a motor mounted on one
side of the inner frame at the attachment point to the outer frame. A
free bearing provides the connection point on the other side.

\begin{figure}
  \begin{center}
  \begin{tabular}{c}
    \includegraphics[width=6cm,angle=270,clip]{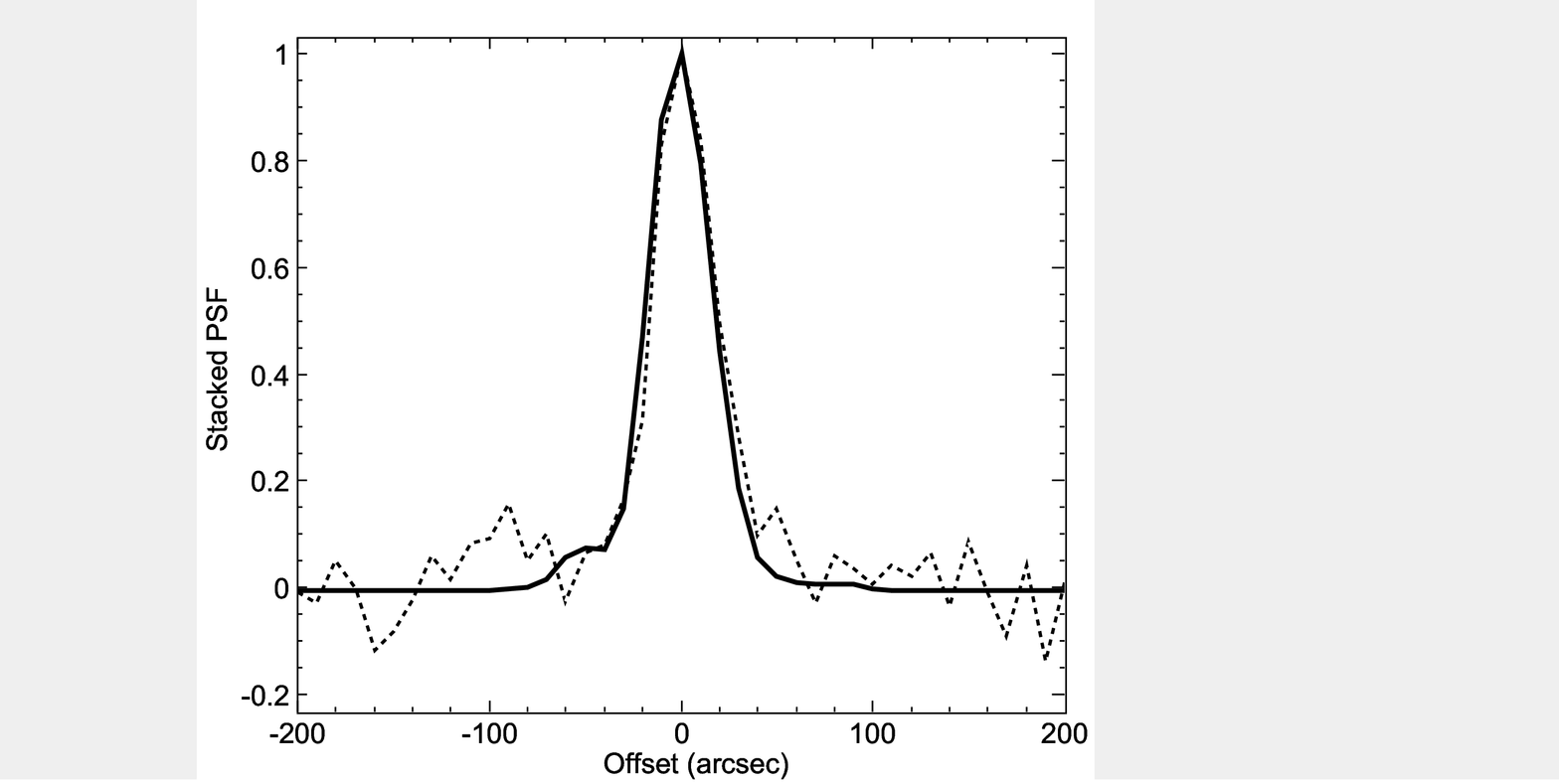}
  \end{tabular}
  \end{center}
  \caption[example]{A cut through the stacked \blast\ 250\,\mic\ flux
    at the positions of VLA 1.4\,GHz radio sources\cite{miller08}
    (dashed line) and through the 250\,\mic\ point spread function
    (solid line). We see that the stack is very well described by the
    PSF, in both position and width. We conclude that our absolute
    pointing is good to $< 2\arcsec$ and that random pointing errors
    are $< 3\arcsec$ rms.
 \label{fig:vlastack}}
   \end{figure}

Pointing is measured in-flight to an accuracy of $\sim 30\arcsec$ rms
by a variety of fine and coarse sensors, including fiber optic
gyroscopes, optical star cameras, a differential GPS, magnetometer and
Sun sensor.  Post-flight pointing reconstruction uses only the
gyroscopes and day-time star trackers~\cite{rex06}. The algorithm is
based on a similar multiplicative extended Kalman
filter\cite{markley03} technique used by \wmap\cite{harman05}, and
modified\cite{pittelkau01} to allow the evaluation of the star
trackers and gyroscopes alignment parameters. The offset between the
star cameras and the submm telescope was measured by repeated
observations of pointing calibrators throughout the flight. We find
that the relative pointing between the star cameras and submm
telescope varies as a function of IF elevation. We apply an
elevation-dependent correction to pitch and yaw with scales or
$\sim$125\arcsec\ and $\sim$20\arcsec, respectively. Post flight
pointing accuracy is verified by a stacking analysis on one of the
extragalactic maps. Using the deep radio E-CDF-S VLA survey at $1.4$
GHz\cite{miller08}, we stack patches of the \blast\ maps centered at
the radio source coordinates, simply summing the flux pixel by pixel
(see Fig.~\ref{fig:vlastack}). We find that the peak in the stacked
map is located within 2\arcsec\ from the nominal position of the
catalogue, indicating that the absolute pointing accuracy is at least
$15$ times smaller than the beam size. Moreover, assuming random
Gaussian pointing errors, we superimpose the synthetic scaled Point
Spread Function (PSF) to the stacked map and convolve it with a
Gaussian profile, modeling the broadening of the PSF due to pointing
jitter. By varying the jitter width, we compute the $\chi^2$ of the
convolved PSF over the stacked data. In this way we estimate the upper
limit in potential random pointing errors to be 3\arcsec.

\acknowledgments     
 
The \blast\ collaboration acknowledges the support of NASA through
grant numbers NAG5-12785, NAG5-13301 and NNGO-6GI11G, the Canadian
Space Agency (CSA), the Science and Technology Facilities Council
(STFC), Canada's Natural Sciences and Engineering Research Council
(NSERC), the Canada Foundation for Innovation, the Ontario Innovation
Trust, the Puerto Rico Space Grant Consortium, the Fondo Istitucional
para la Investigacion of the University of Puerto Rico, and the
National Science Foundation Office of Polar Programs;
C.~B. Netterfield also acknowledges support from the Canadian
Institute for Advanced Research.  L.~Olmi would like to acknowledge
Pietro Bolli for his help with Physical Optics simulations during the
testing phase of BLAST06.  We would also like to
thank the Columbia Scientific Balloon Facility (CSBF) staff for their
outstanding work, the Precision Machining Group at Lawrence Livermore
Laboratory, the support received from Empire Dynamic Structures in the
design and construction of the gondola, Daniele Mortari for helpful
discussions in the development of the Pyramid code, Dan Swetz for
buliding the Fourier transform spectrometer, and Luke Bruneaux, Kyle
Lepage, Danica Marsden, Vjera Miovic, and James Watt for their
contribution to the project.


\bibliography{refs}   

\begin{thebibliography}{10}

\bibitem{grif03}
{Griffin}, M.~J., {Swinyard}, B.~M., and {Vigroux}, L.~G., ``{SPIRE -
  Herschel's Submillimetre Camera and Spectrometer},'' in [{\em IR Space
  Telescopes and Instruments. Edited by John C. Mather . Proceedings of the
  SPIE}{\nolinebreak\hspace{0.1em}]},   {\bf 4850},  686--697 (Mar. 2003).

\bibitem{dev04}
{Devlin}, M.~J., {Ade}, P.~A.~R., {Aretxaga}, I., {Bock}, J.~J., {Chung}, J.,
  {Chapin}, E., {Dicker}, S.~R., {Griffin}, M., {Gundersen}, J., {Halpern}, M.,
  {Hargrave}, P., {Hughes}, D., {Klein}, J., {Marsden}, G., {Martin}, P.,
  {Mauskopf}, P.~D., {Netterfield}, B., {Olmi}, L., {Pascale}, E., {Rex}, M.,
  {Scott}, D., {Semisch}, C., {Truch}, M., {Tucker}, C., {Tucker}, G.,
  {Turner}, A.~D., and {Weibe}, D., ``{The Balloon-borne Large Aperture
  Submillimeter Telescope (BLAST)},'' in [{\em Astronomical Structures and
  Mechanisms Technology. Edited by Antebi, Joseph; Lemke, Dietrich. Proceedings
  of the SPIE, Volume 5498, pp. 42-54 (2004).}{\nolinebreak\hspace{0.1em}]},
  42--54 (Oct. 2004).

\bibitem{hughes02}
{Hughes}, D.~H., {Aretxaga}, I., {Chapin}, E.~L., {Gazta{\~ n}aga}, E.,
  {Dunlop}, J.~S., {Devlin}, M.~J., {Halpern}, M., {Gundersen}, J., {Klein},
  J., {Netterfield}, C.~B., {Olmi}, L., {Scott}, D., and {Tucker}, G.,
  ``{Breaking the `redshift deadlock'- I. Constraining the star formation
  history of galaxies with submillimetre photometric redshifts},'' {\em
  \mnras}~{\bf 335},  871--882 (Oct. 2002).

\bibitem{truch07}
{Truch}, M.~D.~P.and~{Ade}, P.~A.~R., {Bock}, J.~J., {Chapin}, E.~L., {Chung},
  J., {Devlin}, M.~J., {Dicker}, S., {Griffin}, M., {Gundersen}, J.~O.,
  {Halpern}, M., {Hargrave}, P.~C., {Hughes}, D.~H., {Klein}, J., {MacTavish},
  C.~J., {Marsden}, G., {Martin}, P.~G., {Martin}, T.~G., {Mauskopf}, P.,
  {Netterfield}, C.~B., {Olmi}, L., {Pascale}, E., {Patanchon}, G., {Rex}, M.,
  {Scott}, D., {Semisch}, C., {Thomas}, N., {Tucker}, C., {Tucker}, G.~S.,
  {Viero}, M.~P., and {Wiebe}, D.~V., ``{The Balloon-Borne Large Aperture
  Submillimeter Telescope (BLAST) 2005: Calibration and Targeted Sources},''
  {\em ApJ}~{\bf accepted for publication} (Mar. 2008).

\bibitem{pascale07}
{Pascale}, E., {Ade}, P.~A.~R., {Bock}, J.~J., {Chapin}, E.~L., {Chung}, J.,
  {Devlin}, M.~J., {Dicker}, S., {Griffin}, M., {Gundersen}, J.~O., {Halpern},
  M., {Hargrave}, P.~C., {Hughes}, D.~H., {Klein}, J., {MacTavish}, C.~J.,
  {Marsden}, G., {Martin}, P.~G., {Martin}, T.~G., {Mauskopf}, P.,
  {Netterfield}, C.~B., {Olmi}, L., {Patanchon}, G., {Rex}, M., {Scott}, D.,
  {Semisch}, C., {Thomas}, N., {Truch}, M.~D.~P., {Tucker}, C., {Tucker},
  G.~S., {Viero}, M.~P., and {Wiebe}, D.~V., ``{The Balloon-borne Large
  Aperture Submillimeter Telescope: BLAST},'' {\em ApJ}~{\bf accepted for
  publication} (Mar. 2008).

\bibitem{patanchon07}
{Patanchon}, G., {Ade}, P.~A.~R., {Bock}, J.~J., {Chapin}, E.~L., {Devlin},
  M.~J., {Dicker}, S., {Griffin}, M., {Gundersen}, J.~O., {Halpern}, M.,
  {Hargrave}, P.~C., {Hughes}, D.~H., {Klein}, J., {Marsden}, G., {Martin},
  P.~G., {Mauskopf}, P., {Netterfield}, C.~B., {Olmi}, L., {Pascale}, E.,
  {Rex}, M., {Scott}, D., {Semisch}, C., {Truch}, M.~D.~P., {Tucker}, C.,
  {Tucker}, G.~S., {Viero}, M.~P., and {Wiebe}, D.~V., ``{SANEPIC: A Map-Making
  Method for Timestream Data From Large Arrays},'' {\em ApJ}~{\bf accepted for
  publication} (Nov. 2007).

\bibitem{chapin07}
{Chapin}, E.~L., {Ade}, P.~A.~R., {Bock}, J.~J., {Brunt}, C., {Devlin}, M.~J.,
  {Dicker}, S., {Griffin}, M., {Gundersen}, J.~O., {Halpern}, M., {Hargrave},
  P.~C., {Hughes}, D.~H., {Klein}, J., {Marsden}, G., {Martin}, P.~G.,
  {Mauskopf}, P., {Netterfield}, C.~B., {Olmi}, L., {Pascale}, E., {Patanchon},
  G., {Rex}, M., {Scott}, D., {Semisch}, C., {Truch}, M.~D.~P., {Tucker}, C.,
  {Tucker}, G.~S., {Viero}, M.~P., and {Wiebe}, D.~V., ``{The Balloon-borne
  Large Aperture Submillimeter Telescope (BLAST) 2005: A 4 sq. deg Galactic
  Plane Survey in Vulpecula (l=59)},'' {\em ApJ}~{\bf accepted for publication}
  (Oct. 2007).

\bibitem{mckee07}
{McKee}, C.~F. and {Ostriker}, E.~C., ``{Theory of Star Formation},'' {\em
  \araa}~{\bf 45},  565--687 (Sept. 2007).

\bibitem{crutch04}
{Crutcher}, R.~M., {Nutter}, D.~J., {Ward-Thompson}, D., and {Kirk}, J.~M.,
  ``{SCUBA Polarization Measurements of the Magnetic Field Strengths in the
  L183, L1544, and L43 Prestellar Cores},'' {\em \apj}~{\bf 600},  279--285
  (Jan. 2004).

\bibitem{whittet08}
{Whittet}, D.~C.~B., {Hough}, J.~H., {Lazarian}, A., and {Hoang}, T., ``{The
  Efficiency of Grain Alignment in Dense Interstellar Clouds: a Reassessment of
  Constraints from Near-Infrared Polarization},'' {\em \apj}~{\bf 674},
  304--315 (Feb. 2008).

\bibitem{hildebrand00}
{Hildebrand}, R.~H., {Davidson}, J.~A., {Dotson}, J.~L., {Dowell}, C.~D.,
  {Novak}, G., and {Vaillancourt}, J.~E., ``{A Primer on Far-Infrared
  Polarimetry},'' {\em \pasp}~{\bf 112},  1215--1235 (Sept. 2000).

\bibitem{ward00}
{Ward-Thompson}, D., {Kirk}, J.~M., {Crutcher}, R.~M., {Greaves}, J.~S.,
  {Holland}, W.~S., and {Andr{\'e}}, P., ``{First Observations of the Magnetic
  Field Geometry in Prestellar Cores},'' {\em \apjl}~{\bf 537},  L135--L138
  (July 2000).

\bibitem{ostriker01}
{Ostriker}, E.~C., {Stone}, J.~M., and {Gammie}, C.~F., ``{Density, Velocity,
  and Magnetic Field Structure in Turbulent Molecular Cloud Models},'' {\em
  \apj}~{\bf 546},  980--1005 (Jan. 2001).

\bibitem{li06}
{Li}, H., {Griffin}, G.~S., {Krejny}, M., {Novak}, G., {Loewenstein}, R.~F.,
  {Newcomb}, M.~G., {Calisse}, P.~G., and {Chuss}, D.~T., ``{Results of SPARO
  2003: Mapping Magnetic Fields in Giant Molecular Clouds},'' {\em \apj}~{\bf
  648},  340--354 (Sept. 2006).

\bibitem{jones02}
{Jones}, C.~E. and {Basu}, S., ``{The Intrinsic Shapes of Molecular Cloud
  Fragments over a Range of Length Scales},'' {\em \apj}~{\bf 569},  280--287
  (Apr. 2002).

\bibitem{mous99}
{Mouschovias}, T.~C. and {Ciolek}, G.~E., ``{Magnetic Fields and Star
  Formation: A Theory Reaching Adulthood},'' in [{\em NATO ASIC Proc. 540: The
  Origin of Stars and Planetary Systems}{\nolinebreak\hspace{0.1em}]},  {Lada},
  C.~J. and {Kylafis}, N.~D., eds.,  305--+ (1999).

\bibitem{allen03}
{Allen}, A., {Li}, Z.-Y., and {Shu}, F.~H., ``{Collapse of Magnetized Singular
  Isothermal Toroids. II. Rotation and Magnetic Braking},'' {\em \apj}~{\bf
  599},  363--379 (Dec. 2003).

\bibitem{heyer08}
{Heyer}, M., {Gong}, H., {Ostriker}, E., and {Brunt}, C., ``{Magnetically
  Aligned Velocity Anisotropy in the Taurus Molecular Cloud},'' {\em ArXiv
  e-prints}~{\bf 802} (Feb. 2008).

\bibitem{goldsmith08}
{Goldsmith}, P.~F., {Heyer}, M., {Narayanan}, G., {Snell}, R., {Li}, D., and
  {Brunt}, C., ``{Large-Scale Structure of the Molecular Gas in Taurus Revealed
  by High Linear Dynamic Range Spectral Line Mapping},'' {\em ArXiv
  e-prints}~{\bf 802} (Feb. 2008).

\bibitem{goodman90}
{Goodman}, A.~A., {Bastien}, P., {Menard}, F., and {Myers}, P.~C., ``{Optical
  polarization maps of star-forming regions in Perseus, Taurus, and
  Ophiuchus},'' {\em \apj}~{\bf 359},  363--377 (Aug. 1990).

\bibitem{cho05}
{Cho}, J. and {Lazarian}, A., ``{Grain Alignment by Radiation in Dark Clouds
  and Cores},'' {\em \apj}~{\bf 631},  361--370 (Sept. 2005).

\bibitem{chandra53}
{Chandrasekhar}, S. and {Fermi}, E., ``{Magnetic Fields in Spiral Arm},'' {\em
  \apj}~{\bf 118},  113 (July 1953).

\bibitem{zweibel90}
{Zweibel}, E.~G., ``{Magnetic field-line tangling and polarization measurements
  in clumpy molecular gas},'' {\em \apj}~{\bf 362},  545--550 (Oct. 1990).

\bibitem{pelkonen07}
{Pelkonen}, V.-M., {Juvela}, M., and {Padoan}, P., ``{Simulations of polarized
  dust emission},'' {\em \aap}~{\bf 461},  551--564 (Jan. 2007).

\bibitem{falceta08}
{Falceta-Goncalves}, D., {Lazarian}, A., and {Kowal}, G., ``{Studies of regular
  and random magnetic fields in the ISM: statistics of polarization vectors and
  the Chandrasekhar-Fermi technique},'' {\em ArXiv e-prints}~{\bf 801} (Jan.
  2008).

\bibitem{bethell07}
{Bethell}, T.~J., {Chepurnov}, A., {Lazarian}, A., and {Kim}, J.,
  ``{Polarization of Dust Emission in Clumpy Molecular Clouds and Cores},''
  {\em \apj}~{\bf 663},  1055--1068 (July 2007).

\bibitem{vaillancourt08}
{Vaillancourt}, J.~E., {Dowell}, C.~D., {Hildebrand}, R.~H., {Kirby}, L.,
  {Krejny}, M.~M., {Li}, H.-b., {Novak}, G., {Houde}, M., {Shinnaga}, H., and
  {Attard}, M., ``{New Results on the Submillimeter Polarization Spectrum of
  the Orion Molecular Cloud},'' {\em \apjl}~{\bf 679},  L25--L28 (May 2008).

\bibitem{hu97}
{Hu}, W. and {White}, M., ``{A CMB polarization primer},'' {\em New
  Astronomy}~{\bf 2},  323--344 (Oct. 1997).

\bibitem{kamionkowski97}
Kamionkowski, M., Kosowsky, A., and Stebbins, A., ``Statistics of cosmic
  microwave background polarization,'' {\em Phys. Rev. D}~{\bf 55},  7368--7388
  (Jun 1997).

\bibitem{zaldarriaga97}
Zaldarriaga, M. and Seljak, U. c.~v., ``All-sky analysis of polarization in the
  microwave background,'' {\em Phys. Rev. D}~{\bf 55},  1830--1840 (Feb 1997).

\bibitem{page07}
{Page}, L., {Hinshaw}, G., {Komatsu}, E., {Nolta}, M.~R., {Spergel}, D.~N.,
  {Bennett}, C.~L., {Barnes}, C., {Bean}, R., {Dor{\'e}}, O., {Dunkley}, J.,
  {Halpern}, M., {Hill}, R.~S., {Jarosik}, N., {Kogut}, A., {Limon}, M.,
  {Meyer}, S.~S., {Odegard}, N., {Peiris}, H.~V., {Tucker}, G.~S., {Verde}, L.,
  {Weiland}, J.~L., {Wollack}, E., and {Wright}, E.~L., ``{Three-Year Wilkinson
  Microwave Anisotropy Probe (WMAP) Observations: Polarization Analysis},''
  {\em \apjs}~{\bf 170},  335--376 (June 2007).

\bibitem{ben04}
{Beno{\^ i}t}, A., {Ade}, P., {Amblard}, A., {Ansari}, R., {Aubourg}, {\' E}.,
  {Bargot}, S., {Bartlett}, J.~G., {Bernard}, J.-P., {Bhatia}, R.~S.,
  {Blanchard}, A., {Bock}, J.~J., {Boscaleri}, A., {Bouchet}, F.~R.,
  {Bourrachot}, A., {Camus}, P., {Couchot}, F., {de Bernardis}, P.,
  {Delabrouille}, J., {D{\' e}sert}, F.-X., {Dor{\' e}}, O., {Douspis}, M.,
  {Dumoulin}, L., {Dupac}, X., {Filliatre}, P., {Fosalba}, P., {Ganga}, K.,
  {Gannaway}, F., {Gautier}, B., {Giard}, M., {Giraud-H{\' e}raud}, Y.,
  {Gispert}, R., {Guglielmi}, L., {Hamilton}, J.-C., {Hanany}, S.,
  {Henrot-Versill{\' e}}, S., {Kaplan}, J., {Lagache}, G., {Lamarre}, J.-M.,
  {Lange}, A.~E., {Mac{\'{\i}}as-P{\' e}rez}, J.~F., {Madet}, K., {Maffei}, B.,
  {Magneville}, C., {Marrone}, D.~P., {Masi}, S., {Mayet}, F., {Murphy}, A.,
  {Naraghi}, F., {Nati}, F., {Patanchon}, G., {Perrin}, G., {Piat}, M.,
  {Ponthieu}, N., {Prunet}, S., {Puget}, J.-L., {Renault}, C., {Rosset}, C.,
  {Santos}, D., {Starobinsky}, A., {Strukov}, I., {Sudiwala}, R.~V.,
  {Teyssier}, R., {Tristram}, M., {Tucker}, C., {Vanel}, J.-C., {Vibert}, D.,
  {Wakui}, E., and {Yvon}, D., ``{First detection of polarization of the
  submillimetre diffuse galactic dust emission by Archeops},'' {\em \aap}~{\bf
  424},  571--582 (Sept. 2004).

\bibitem{fink99}
{Finkbeiner}, D.~P., {Davis}, M., and {Schlegel}, D.~J., ``{Extrapolation of
  Galactic Dust Emission at 100 Microns to Cosmic Microwave Background
  Radiation Frequencies Using FIRAS},'' {\em \apj}~{\bf 524},  867--886 (Oct.
  1999).

\bibitem{Olmi02}
{Olmi}, L., ``{Optical designs for submillimeter-wave spherical-primary
  (sub)orbital telescopes and novel optimization techniques},'' in [{\em Highly
  Innovative Space Telescope Concepts Edited by Howard A. MacEwen. Proceedings
  of the SPIE}{\nolinebreak\hspace{0.1em}]},   {\bf 4849},  245--256 (Dec.
  2002).

\bibitem{bock98}
{Glenn}, J., {Bock}, J.~J., {Chattopadhyay}, G., {Edgington}, S.~F., {Lange},
  A.~E., {Zmuidzinas}, J., {Mauskopf}, P.~D., {Rownd}, B., {Yuen}, L., and
  {Ade}, P.~A., ``{Bolocam: a millimeter-wave bolometric camera},'' in [{\em
  Proc. SPIE, Advanced Technology MMW, Radio, and Terahertz Telescopes, Thomas
  G. Phillips; Ed.}{\nolinebreak\hspace{0.1em}]},   {\bf 3357},  326--334 (July
  1998).

\bibitem{row03}
{Rownd}, B., {Bock}, J.~J., {Chattopadhyay}, G., {Glenn}, J., and {Griffin},
  M.~J., ``{Design and performance of feedhorn-coupled bolometer arrays for
  SPIRE},'' in [{\em Millimeter and Submillimeter Detectors for Astronomy.
  Edited by Phillips, Thomas G.; Zmuidzinas, Jonas. Proceedings of the
  SPIE}{\nolinebreak\hspace{0.1em}]},   {\bf 4855},  510--519 (Feb. 2003).

\bibitem{renb04}
{Renbarger}, T., {Chuss}, D.~T., {Dotson}, J.~L., {Griffin}, G.~S., {Hanna},
  J.~L., {Loewenstein}, R.~F., {Malhotra}, P.~S., {Marshall}, J.~L., {Novak},
  G., and {Pernic}, R.~J., ``{Early Results from SPARO: Instrument
  Characterization and Polarimetry of NGC 6334},'' {\em \pasp}~{\bf 116},
  415--424 (May 2004).

\bibitem{masi06}
{Masi}, S., {Ade}, P.~A.~R., {Bock}, J.~J., {Bond}, J.~R., {Borrill}, J.,
  {Boscaleri}, A., {Cabella}, P., {Contaldi}, C.~R., {Crill}, B.~P., {de
  Bernardis}, P., {de Gasperis}, G., {de Oliveira-Costa}, A., {de Troia}, G.,
  {di Stefano}, G., {Ehlers}, P., {Hivon}, E., {Hristov}, V., {Iacoangeli}, A.,
  {Jaffe}, A.~H., {Jones}, W.~C., {Kisner}, T.~S., {Lange}, A.~E., {MacTavish},
  C.~J., {Marini Bettolo}, C., {Mason}, P., {Mauskopf}, P.~D., {Montroy},
  T.~E., {Nati}, F., {Nati}, L., {Natoli}, P., {Netterfield}, C.~B., {Pascale},
  E., {Piacentini}, F., {Pogosyan}, D., {Polenta}, G., {Prunet}, S.,
  {Ricciardi}, S., {Romeo}, G., {Ruhl}, J.~E., {Santini}, P., {Tegmark}, M.,
  {Torbet}, E., {Veneziani}, M., and {Vittorio}, N., ``{Instrument, method,
  brightness, and polarization maps from the 2003 flight of BOOMERanG},'' {\em
  \aap}~{\bf 458},  687--716 (Nov. 2006).

\bibitem{rex06}
{Rex}, M., {Chapin}, E., {Devlin}, M.~J., {Gundersen}, J., {Klein}, J.,
  {Pascale}, E., and {Wiebe}, D., ``{BLAST autonomous daytime star cameras},''
  in [{\em Ground-based and Airborne Instrumentation for Astronomy. Edited by
  McLean, Ian S.; Iye, Masanori. Proceedings of the SPIE, Volume 6269, pp.
  62693H (2006).}{\nolinebreak\hspace{0.1em}]},  {\em Presented at the Society
  of Photo-Optical Instrumentation Engineers (SPIE) Conference} {\bf 6269}
  (July 2006).

\bibitem{markley03}
{Markley}, F.~L., ``{Attitude Error Representations for Kalman Filtering},''
  {\em Journal of Guidance, Control, and Dynamics}~{\bf 26},  311--317 (2003).

\bibitem{harman05}
Harman, R.~R., ``{Wilkinson Microwave Anisotropy Probe (WMAP) Attitude
  Estimation Filter Comparison},'' tech. rep., Flight Mechanics Symposium;
  Greenbelt, MD (2005).

\bibitem{pittelkau01}
{Pittelkau}, M.~E., ``{Kalman Filtering for Spacecraft System Alignment
  Calibration},'' {\em Journal of Guidance, Control, and Dynamics}~{\bf 24},
  1187--1195 (2001).

\bibitem{miller08}
{Miller}, N.~A., {Fomalont}, E.~B., {Kellermann}, K.~I., {Mainieri}, V.,
  {Norman}, C., {Padovani}, P., {Rosati}, P., and {Tozzi}, P., ``{The VLA
  1.4GHz Survey of the Extended Chandra Deep Field South: First Data
  Release},'' {\em ArXiv e-prints}~{\bf 804} (Apr. 2008).

\end{thebibliography}
\bibliographystyle{spiebib}   

\end{document}